\documentclass[sigplan,screen]{acmart}

\setcopyright{none}
\renewcommand\footnotetextcopyrightpermission[1]{} 

\begin{document}

\title{SeDe: Balancing Blockchain Privacy and Regulatory Compliance By Selective De-Anonymization}

\author{Amit Chaudhary}
\affiliation{\country{}}
\email{amit.chaudhary.3@warwick.ac.uk}

\author{Hamish Ivey-Law }
\affiliation{\country{}}
\email{hamish@ivey-law.name}

\thispagestyle{plain}
\pagestyle{plain}

\begin{abstract}

Privacy is one of the essential pillars for the widespread adoption of blockchains, but public blockchains are transparent by nature. Modern analytics techniques can easily subdue the pseudonymity feature of a blockchain user. Some applications have been able to provide practical privacy protections using privacy-preserving cryptography techniques. However, malicious actors have abused them illicitly, discouraging honest actors from using privacy-preserving applications as "mixing" user interactions and funds with anonymous bad actors, causing compliance and regulatory concerns. 

In this paper, we propose a framework that balances privacy-preserving features by establishing a regulatory and compliant framework called Selective De-Anonymization (SeDe). The adoption of this framework allows privacy-preserving applications on blockchains to de-anonymize illicit transactions by recursive traversal of subgraphs of linked transactions. Our technique achieves this without leaving de-anonymization decisions or control in the hands of a single entity but distributing it among multiple entities while holding them accountable for their respective actions. To instantiate, our framework uses threshold encryption schemes and Zero-Knowledge Proofs (ZKPs).

\end{abstract}

\maketitle

\section{Introduction}

The most popular blockchains, Bitcoin and Ethereum, are public.
The transparent feature of blockchain allows anyone to access, edit, and audit the transactions on the blockchain. The only guarantee of user privacy comes from pseudonymity, which is overcome by advanced analytics techniques that can precisely identify users by generating user profiles by linking the historical interactions and even name ids \cite{ens} of users. This transparency of public blockchain seriously threatens users' privacy, leading to frequent malicious attacks by bad actors. There have been attempts to provide privacy-preserving mechanisms for blockchain using zero-knowledge proofs \cite{zcash,tornado}. However, applications incorporating these mechanisms for privacy have been lucrative to criminals who exploit this very private nature to launder illicitly possessed assets by evading observation from regulatory bodies \cite{tornado-sanction}.

In an ideal scenario, users should be able to conduct private transactions while remaining compliant and avoid associating with malicious actors. To address this, some solutions have been briefly discussed as an idea for a possible route for compliance in previous works like \textbf{Selective De-Anonymization} \cite{a16z}. The voluntary selective de-anonymization allows users to render view-only access to their transactions to authorities in an act of compliance. However, it relies on the assumption that a malicious actor will choose to give this view-only access. This makes such voluntary disclosures less appealing.

This paper tries to solve the privacy and regulatory compliance dilemma by \textbf{Involuntary Selective De-Anonymization}, or \textbf{SeDe}, in privacy-preserving on-chain applications by using de-anonymization techniques even when the user is not volunteering to cooperate for compliance purposes \cite{mutal, papr, yoso, acc-privacy}. We show a practical implementation of involuntary selective de-anonymization in privacy-preserving on-chain protocols where privacy is achieved with compliance, allowing the tracing of illicit asset transactions. We distribute the de-anonymization ability to multiple parties to reduce the chances of power abuse while keeping such parties accountable for their actions. 

In our solution, users' privacy is protected because of strong guarantees against the collusion of any privacy revoker using multi-party computation schemes. Simultaneously, the compliance features enable the publicly verified request of the privacy revoker to be propagated and executed to reveal suspect transactions only when a sufficient number of independent participants agree to the privacy revocation request. 

\section{Problem Statement}

The privacy-preserving protocols on blockchains achieve privacy property by pooling together a particular asset in a single "pool" smart contract from different sources (e.g. wallets like MetaMask\footnote{https://metamask.io/}). This pool contract can be considered a container for all users' private or shielded accounts and deposit transactions as funding a shielded account. However, unlike regular Ethereum accounts, the balance of a shielded account is not public. An external observer may only see the aggregated balance of the pool contract. When the number of such shielded accounts grows as various distinct sources deposit into the pool, it becomes very difficult to identify the user who performs any future transaction using assets from its shielded account. See Section \ref{zk-privacy-apps} and \ref{example-nova} for more technical details.

Without any compliance measures, however, such privacy-enabling applications have introduced a severe issue: they have become a go-to place to launder assets from various sources of illicit assets. Malicious actors use it to evade tracking illicit activity by funding their shielded accounts with stolen assets. It allows them to privately move these assets between shielded accounts or withdraw to a regular wallet address in the future. These applications, therefore, pose the risk of mixing legitimate funds with potential criminal or illicitly achieved funds. The non-custodial and decentralized nature of these privacy-preserving solutions creates the regulatory and compliance challenge in honest users not adopting the privacy-preserving applications due to the threat of sanctioning and litigations from regulatory bodies and financial institutions. This dilemma of privacy protection and regulatory compliance in blockchain applications deters users from adopting blockchain technology.

Adopting compliance measures from centralized counterparts of such applications is either not simple or only comes with hurting user privacy or other aspects. Nevertheless, there have been some attempts to resolve this privacy protection and compliance dilemma. However, such attempts have only provided partial solutions. We briefly discuss some of these previously attempted solutions below.

\subsection{Imposing Deposit Limit}
A standard measure many applications adopt is limiting how much can be deposited at once or in a specified period. This throttles the inflow of laundering of assets gained from illicit activities. However, some amount still may be laundered, and it causes inconvenience for many users if the imposed limit is too restricting. This limits the scope of application to small peer-to-peer payments, excluding anything beyond that. 

\subsection{Sanctioned Addresses}
The \textit{OFAC} put many addresses in a list of sanctioned entities list (\textit{SDN}) which were linked to illicit transactions on Tornado. Blockchain analysis firms like Chainalysis\footnote{https://go.chainalysis.com/chainalysis-oracle-docs.html} provide oracles to access this list in smart contracts, which many applications use as a blocklist in their smart contracts.

However, circumventing this measure is not hard because the sanction list is not updated in real time. Any malicious actor could immediately deposit stolen funds before the list is updated or even send stolen assets to a fresh new address and deposit from that address. Besides, it is uncertain whether there would be any recovery mechanism for addresses put mistakenly in the on-chain sanction list.

\subsection{Blockchain Analysis Tools}
Multiple blockchain analytic firms like Chainalysis\footnote{https://www.chainalysis.com/} and TRM Labs\footnote{https://www.trmlabs.com/} render tools and services for tracking transactions related to illicit funds. Moreover, they have also been successful in tracking\footnote{https://www.trmlabs.com/post/north-koreas-lazarus-group-moves-funds-through-tornado-cash} such transactions.

However, such tools are not accurate because the margin of error is significant in the classification metrics, leading to false negatives that are not tracked or made public to ensure the robust performance of the analysis tools.

\subsection{View-Only Access}
Some privacy-preserving applications propose only a view-access for transactions by having a separate \textit{view key} that could decrypt encrypted transaction data (but not spend it). If required, users may share this key with the authorities for regulatory reasons without giving up the authority to spend it. This is a suitable mechanism only for honest users who are willing to volunteer in the act of compliance. The mechanism lacks to enforce and sound compliance guarantees for malicious users who will not willingly disclose their transactions by voluntary mechanism. 

\subsection{Association Sets}
The most recent proposed solution for achieving compliance is through the use of association sets in privacy pools \cite{privacy-pools}. It lets users associate with only a subset of deemed good deposits by proving the inclusion of good actors and the exclusion of bad actors. This limits the user's anonymity proportional to the size of this subset rather than being equal to all such deposits. Limiting association sets can make it easier for analysis tools to link transactions. There could also be cases where laundering of assets is detected after much delay\footnote{https://cointelegraph.com/news/the-aftermath-of-axie-infinity-s-650m-ronin-bridge-hack}. By then; many honest users might get associate with the illicit transaction. Moreover, there is no compliance mechanism in case a malicious actor successfully makes a private withdrawal because of the lack of de-anonymization.

\section{Overview Of Selective De-Anonymization} \label{overview}

We present SeDe in the context of privacy-preserving applications that live on public blockchain as a set of smart contracts. Such applications adopt a similar architecture to that of the ZCash \cite{zcash}. The transactions in these kinds of applications simulate the spending of currency notes, also sometimes referred to as a UTXO (Unspent Transaction Output). Performing a transaction in such applications is equivalent to spending a number of notes, the same as spending physical notes in real life. The so-called "nullifiers" of each note are constrained to be revealed when spending them. Nullifiers are elements that are unique and cryptographically bound to a note. The spent notes' nullifiers are marked in the application to prevent notes with already marked nullifiers from being spent again. A fresh set of new notes is created such that the total value is conserved in the state of application. The bearer of the newly created notes, however, may change in transaction. In the case of an external deposit (spending 0 notes), an equivalent amount of notes is created with the depositor as bearer. Except initial deposit, the user performing the transaction does not reveal any sensitive details like the identities of the sender and receiver or asset amounts in any public domain including the application smart contracts. Rather, it provides a zero-knowledge proof of correct computation of transaction, which the verifier within the application verifies. See Sec. \ref{example-nova} for a technical example of such an application.

SeDe is based on the core idea of being able to trace and de-anonymize a subgraph of linked illicit transactions by following the spending of notes created by executing the transaction. De-anonymizing a transaction $T$ in the context of the application described earlier is equivalent to getting access to information about notes $N_j$ that were created in that transaction since a note contains sensitive fields like owner $id$ and its value. 

Before, describing the technique we define the following actors involved in the process below.

\begin{itemize}
    \item \textbf{User}, $\mathcal{U}$: The user of the privacy-preserving application. A user can be honest whose sole purpose is achieving privacy in its transactions. It can also be a malicious actor who may attempt to launder assets by exploiting the very private property.
    \item \textbf{Key Issuer}, $\mathcal{I}$: Key issuer is an application-specific entity to enroll users as members of the platform by letting them have a unique identity $id$.
    \item \textbf{Guardians}, $\mathcal{G}$: A set of entities or individuals for gate-keeping transaction de-anonymization. De-anonymization is possible only if a subset of $n$ guardians form a quorum of minimum size $t$. Guardians can only dictate the decision of de-anonymization, plaintext transaction is not revealed to them.
    \item \textbf{Revoker}, $\mathcal{R}$: Only a revoker can de-anonymize and see plain transaction data. But it cannot do so unless it sends a publicly verifiable signal of de-anonymization request to guardians who must agree first. 
\end{itemize}

Normally, while constructing a legitimate transaction request, such applications require the user to adhere to a set of constraints to generate a valid ZK proof just for the sake of the privacy of the transaction. We extend this set of constraints that the user will adhere to, in an act of compliance. This also aids later in the de-anonymization of suspected illicit transactions.

In our proposed framework, apart from other details, we constrain the user to include the encryptions of the newly created notes, $N_j$ in the transaction payload. The encryption must be double-layered, first encrypting by revoker's public key $P_\mathcal{R}$ followed by encrypting the result of previous encryption by public key $P_\mathcal{G}$ representing the guardian set. Apart from the resulting ciphertext $E_{\mathcal{G}}(E_{\mathcal{R}}(N_i))$, the user must also include a proof $\boldsymbol{\pi}$ in the payload proving that it performed the encryptions correctly.

\begin{figure}[h]
    \centering
    \includegraphics[scale=0.28]{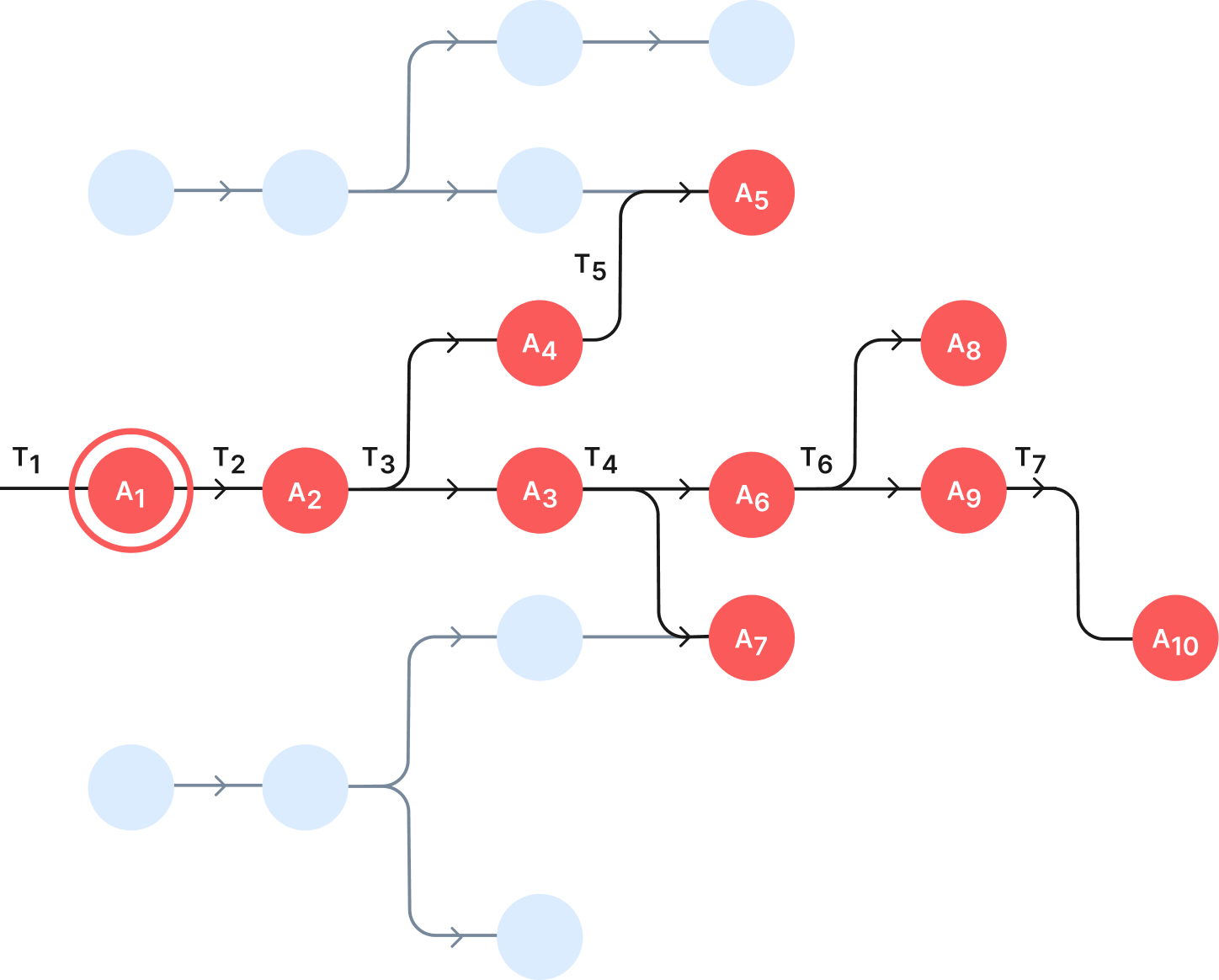}
    \caption{Subgraph of linked illicit transactions. A tainted account, $A_i$ (shown in red) receives illicitly sourced assets by executing some transaction, $T_j$ in which tainted notes were spent.}
    \label{fig:tx-subgraph}
\end{figure}

\begin{figure*}[h]
    \centering
    \includegraphics[scale=0.35]{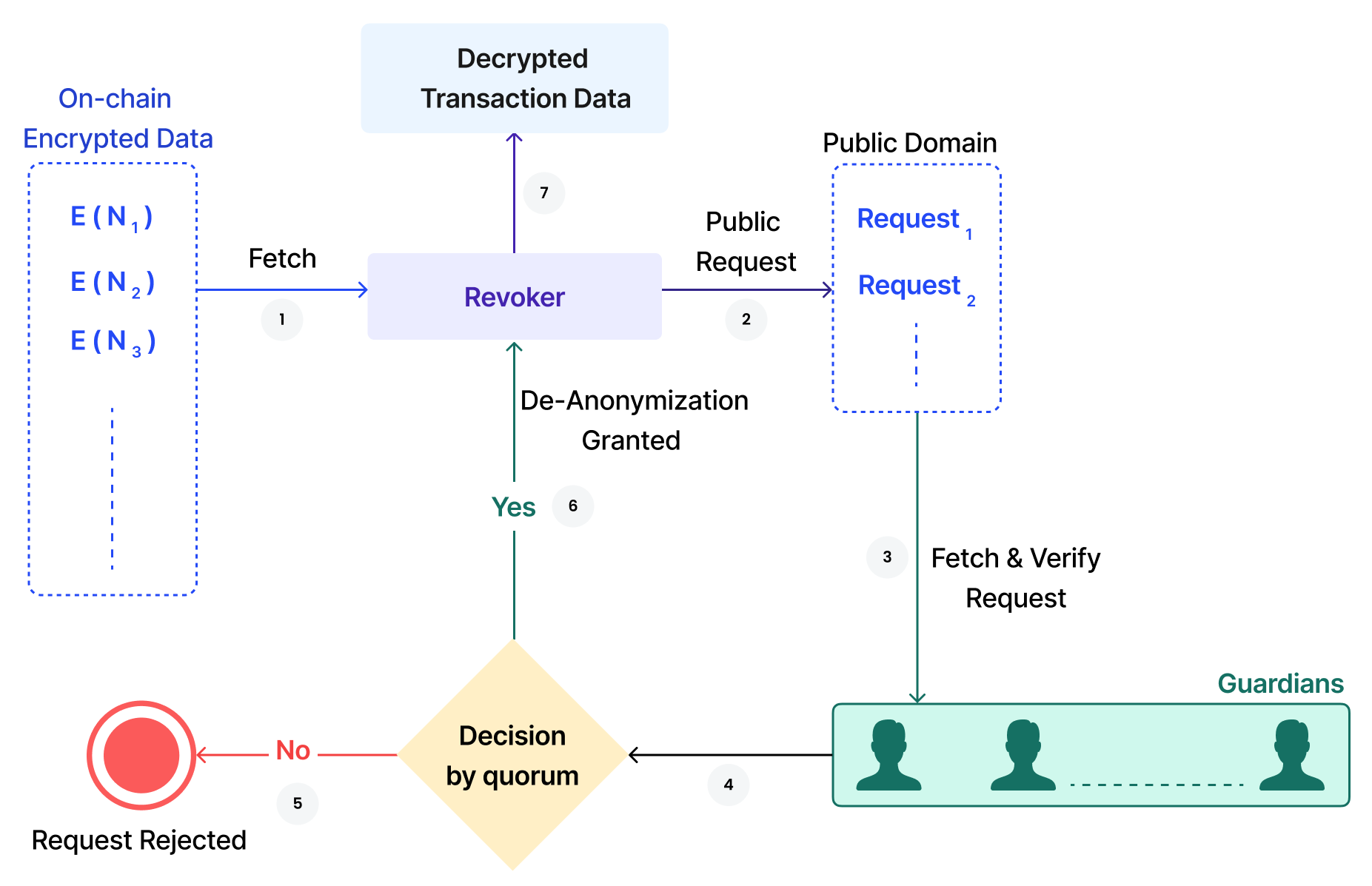}
    \caption{SeDe Overview}
    \label{fig:sede-overview}
\end{figure*}

Figure \ref{fig:tx-subgraph} shows a graph consisting of shielded accounts, $A_i$ as nodes that are connected by transaction $T_i$ edges. A connected pair of accounts represents the flow of asset notes between the accounts. Highlighted nodes, $A_i$ are the accounts that received at least one note sourced from illicit transactions. We refer to these as tainted notes or tainted assets. All such connected nodes form a subgraph of illicit transactions. A bad actor, for laundering such assets, sends an initial deposit transaction $T_1$ to fund a shielded account $A_1$. Since it is an initial deposit i.e. assets are moving from an external source to a pool smart contract, $T_1$ can be identified as illicit, which is what usually happens. Any subsequent transaction like $T_2$ to launder assets further by transferring assets to another shielded account $A_2$ remains private and untraceable by external observers. The launderer may also choose to split tainted asset amounts as shown by executing transaction $T_3$ that funds the accounts $A_3$ and $A_4$ with tainted notes. And so on.

The overview of the de-anonymization of an individual transaction is demonstrated in figure \ref{fig:sede-overview}. In the event of an illicit transaction, the de-anonymization process is undertaken by taking the following steps. (1) It identifies and fetches the encrypted data of transaction a $T$ from on-chain. (2) Revoker then signs a message with its private key $p_{\mathcal{R}}$ representing a de-anonymization request for $T$ and publishes the request to some public domain. (3) Guardians fetch a request from a queue of requests and verify the signature to ensure that the request indeed came from the revoker. (4) Then guardians start the decision-making process and form a quorum to grant permission to the revoker. Then if (5) quorum is not reached, the request is rejected. Or if (6) quorum is reached, permission is granted to the revoker by sending cryptographic contributions from guardians. (7) After permission is granted, the revoker is enabled to decrypt the encrypted transaction data and get information about the tainted notes $N_j$ created in the transaction.

Since any spending of these tainted notes $N_j$ in a future transaction requires the revelation of nullifiers $\eta_j$, the revoker may scan the network for all transactions to find transactions $T^{\prime}$ in which any of $\eta_j$ were revealed. If found, such transactions are identified as illicit and the revoker performs the same operation as above on all such transactions. The process is performed continuously by recursively tracing such transactions until a subgraph of all transactions originating from an illicit source is traversed. 

The technique design adheres to certain properties that discourage involved parties from acting maliciously and enforce accountability.

\begin{itemize}
    \item \textbf{Accountable Anonymity:} User is cryptographically bound to perform transactions in a compliant way such that an attempt to act maliciously leads to losing anonymity to lawful authorities and actions. 
    \item \textbf{Accountable De-Anonymization:} Revoker cannot proceed with the de-anonymization without posting a verifiable message publicly requesting the same. Guardians cannot reveal transaction data themselves without cooperation from revoker even if they collude in any number.
    \item \textbf{Non-Fabrication:} An honest user can prove its exclusion from involvement in the illicit transaction subgraph even if revoker and guardians collude to put false accusations on an honest user.
\end{itemize}

\section{Background}

\subsection{ZKPs in Privacy Preserving Applications} \label{zk-privacy-apps}
Public blockchains usually have a simple account-based model for balance accounting purposes. This is not a favorable design for privacy-focused applications.

Private applications like ZCash blockchain and Tornado Cash adopt a so-called \textit{UTXO (Unspent Transaction Output)} based model which simulates the physical cash notes-like interaction (called JoinSplit), thereby achieving privacy properties of peer-to-peer transactions. For simplicity, a UTXO can be considered equivalent to currency notes. Having several notes allows you to spend it in a transaction for whatever purpose.

The UTXO-based model can achieve privacy in transactions via ZKPs (like zkSNARKs). A Merkle tree data structure is maintained in such private transactions enabling applications or chains. This tree keeps track of all notes to ever exist by holding the hash of each such note, also referred to as note \textbf{commitment}, $c$. To (privately) spend a note, $N$ in a transaction, $T$ the user has to generate a ZK proof, $\boldsymbol{\pi}$ that proves the correct computation of the transaction without revealing any identifying information. Each note is bound to a unique value usually referred to as \textbf{nullifier}. The nullifier, $\eta$ of a note is required to be revealed in the transaction payload when that note is being spent. The application marks each spent note's nullifiers to keep track of already spent (or nullified) notes to prevent any double-spend of the same in the future. After spent notes are nullified (or equivalently destroyed), fresh new notes are created such that the total value is conserved.

\begin{figure}
    \centering
    \includegraphics[scale=0.18]{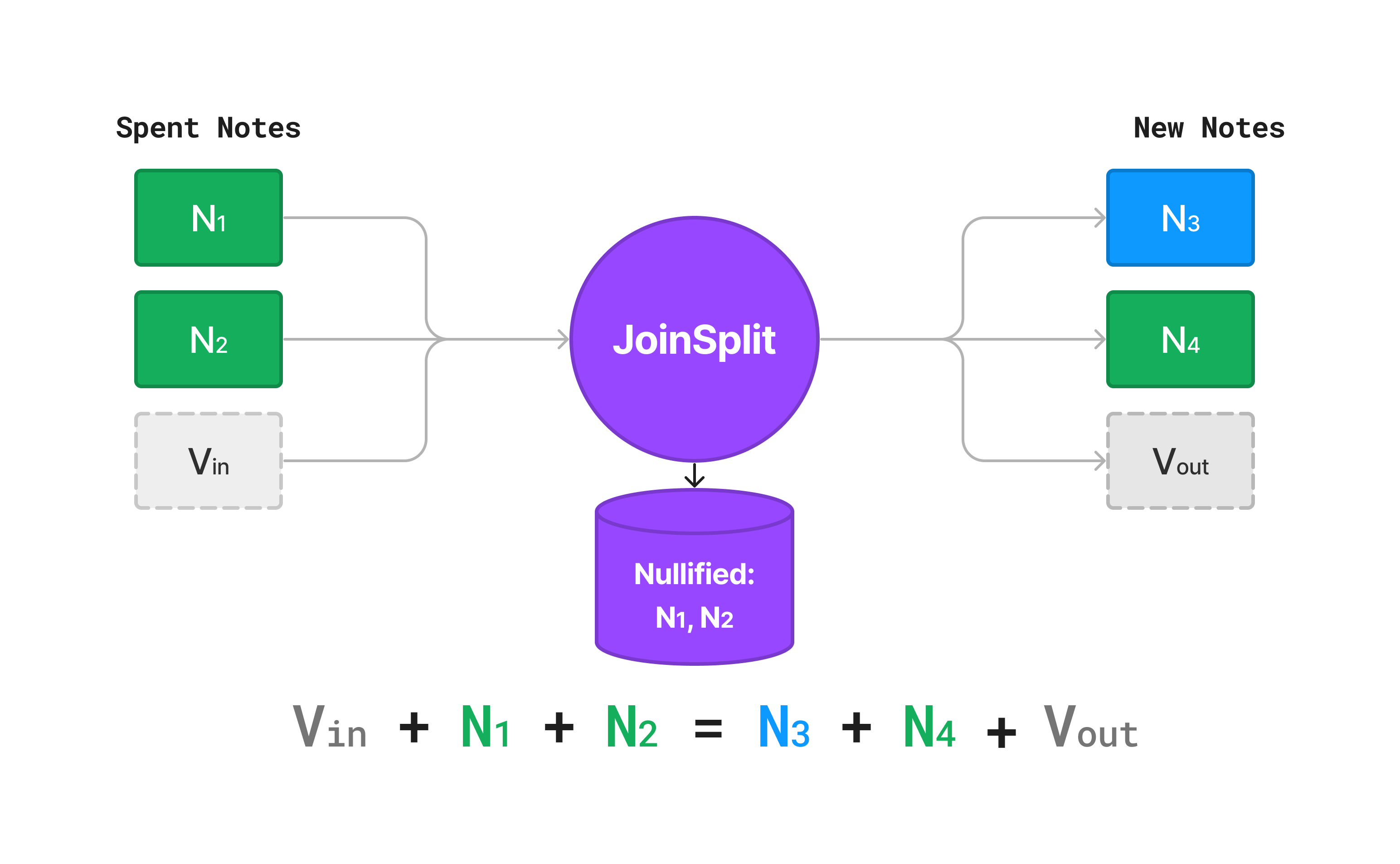}
    \caption{JoinSplit transaction}
    \label{fig:joinsplit}
\end{figure}

The figure \ref{fig:joinsplit} represents notes involved in a JoinSplit transaction. These transactions can be broadly divided into three types. (1) \textbf{Deposit} transaction where a non-zero external amount $V_{in}$ sent with transaction payload and equally valued one or more notes ($N_3$ and $N_4$) are created. Any spent notes' ($N_1$ and $N_2$) value or any external withdraw amount $V_{out}$ are zero. (2) \textbf{Withdraw} transaction where non-zero value notes ($N_1$ and $N_2$) are spent to withdraw some non-zero value $V_{out}$ less than or equal to the sum of values of spent notes. Any change value left is accounted for by the creation of one or more notes ($N_3$ and $N_4$) for the user. (3) \textbf{Transfer} transaction where no external value is revealed (i.e. $V_{in} = V_{out} = 0$). Some notes ($N_1$ and $N_2$) are spent to create some new notes ($N_3$ and $N_4$) such that values remain conserved, However, the bearer of some note ($N_3$ in the figure) may change, which in effect simulates a fully-private peer-to-peer transfer of value.

The transaction payload usually contains the proof $\boldsymbol{\pi}$, commitments of new notes $c_i$, nullifiers of notes being spent, $\eta_i$ and other application-specific public inputs $\rho^{\prime}$.
\begin{equation}
    T \equiv (\boldsymbol{\pi}, c_i, \eta_i, \rho^{\prime})
\end{equation}

An account (also referred to as a shielded account), $A$ is a collection of notes with the same address or key, that has authority to spend those. The balance of an account is simply the sum of all unspent notes in that account. Execution of a transaction $T$ changes the state of one or more such accounts.

\subsection{Example: Tornado Nova} \label{example-nova}
Although multiple ZCash-inspired privacy-preserving applications exist in the space they work roughly the same way. We take an example of Tornado Nova or simply Nova\footnote{https://github.com/tornadocash/tornado-nova}. Nova was a slightly advanced version of Tornado Cash as it allowed an arbitrary amount of peer-to-peer fully private (shielded) transactions. A note, $N$ in Nova was simply a tuple of amount $v$, owner's public key $K$ and a blinding $r_b$ (a random number) i.e.:
\begin{equation}
    N \equiv (v, K, r_b)
\end{equation}

The commitment $c$ of the note is simply the poseidon hash \cite{poseidon} of $N$:
\begin{equation}
    c = H(v, K, r_b)
\end{equation}

The commitments of all notes ever created are stored in a merkle tree data structure maintained in Nova's smart contract.

Nullifier $\eta$ of a note whose commitment $c$ is inserted at index $i$ in the merkle tree is defined in Nova as:
\begin{equation}
    \eta = H(c, i, H(k, c, i))
\end{equation}

where $k$ is the private key corresponding to public key $K$.

In a private transaction on Nova, user is only able to spend notes whose commitments are already inserted in the tree and which are not nullified already.

To spend a note(s) $N_{spend}$ user proves via a ZKP that:
\begin{enumerate}
    \item It knows a tuple $N_{spend}$ such that its commitment $c_{spent}$ is at some index $i_{spend}$ in the tree.
    \item It knows the opening in the tree at the index $i_{spend}$.
    \item It has the knowledge of the private key $k$ corresponding to public key in the tuple $N_{spend}$.
    \item The calculated nullifer $\eta_{spend}$ was calculated correctly.
\end{enumerate}

It also proves statements involving creation of new note(s) $N_{new}$:
\begin{enumerate}
    \item It calculated the commitment of new note $c_{new}$ correctly.
    \item The total value of new notes is equal to that of notes that were spent i.e. total value is conserved.
\end{enumerate}

In a successful transaction, the commitment of a newly created note $c_{new}$ is inserted at the next empty index $i_{new}$ into a Merkle tree that exists as a smart contract on the blockchain. And the nullifiers $\eta_{spend}$ are recorded by the smart contract to prevent double-spend of already spent $N_{spend}$.

\subsection{Compliance Measures}
To prevent using such applications in any illicit activity such as money laundering by exploiting their private nature, most of the applications deployed so far have only adopted simple measures.

Tornado Cash made available a compliance tool\footnote{https://docs.tornadoeth.cash/docs/compliance-tool} for their legitimate users that allowed them to prove the origin of their funds. In other words, it allowed to re-link source address in the deposit and the destination address in the withdrawal, together comprising a report of transaction history.

Others like Aztec's zk.money and Tornado Nova simply put a limit (e.g. 1 or 2 ETH) on the deposit amount going into such applications. This is to prevent huge value, likely coming from a large-scale hack, from flowing into it.

Many protocols including privacy-preserving ones, block an address from interaction with their service if the address was sanctioned by government. Hence, complying with the law. Tornado Cash adopted this measure, however only on their client-facing interface.

A few newer protocols adopted other measures including imposing of KYC for their users or using third-party blockchain analysis tools.

\section{Prelimnaries}
\subsection{Zero-Knowledge Proofs}
Zero-knowledge proof (ZKP) is a cryptographic technique that allows a party, a prover to prove to another party, a verifier that it knows a secret without actually revealing a secret by following a set of complex mathematical operations. ZKPs can be traced back to the early 1980s when a group of researchers first introduced it \cite{zkp}. Though it lacked practicality at the time.

ZKPs must satisfy three properties:
\begin{itemize}
    \item \textbf{Completeness}: If the statement is true, both the prover and verifier are following the protocol, then the verifier will accept the proof.
    \item \textbf{Soundness}: If the statement is false, no prover can convince the verifier that it is true with any significant probability.
    \item \textbf{Zero Knowledge}: If a statement is true, the verifier learns nothing other than the fact that the statement is true.
\end{itemize}

A \textbf{SNARK} (Succinct Non-Interactive Argument of Knowledge) is a particular proof system where proof is small in size (succinct), verifiable in a short time, and does not require any interaction between prover and verifier other than sending the proof itself. If the proof system allows for proof verification without revealing any secret inputs used in constructing proof it is called \textbf{zkSNARK}.

Although some multiple proving systems or algorithms are used to generate ZK proofs, most privacy-preserving applications on blockchains have adopted a particular system called \textbf{Groth16} \cite{groth16}. This is because of the desired properties of the proofs (SNARK) that it facilitates that are efficient for applications on blockchains.

\subsection{Merkle Tree}
Merkle tree is a data structure that is commonly used to verify the integrity of large amounts of data blocks. It is a binary tree data structure where each node has two children and each node is a hash of its children. The root of the tree eventually becomes a culmination of all data blocks whose hashes are leaf nodes.

\begin{figure}[h]
    \centering
    \includegraphics[scale=0.15]{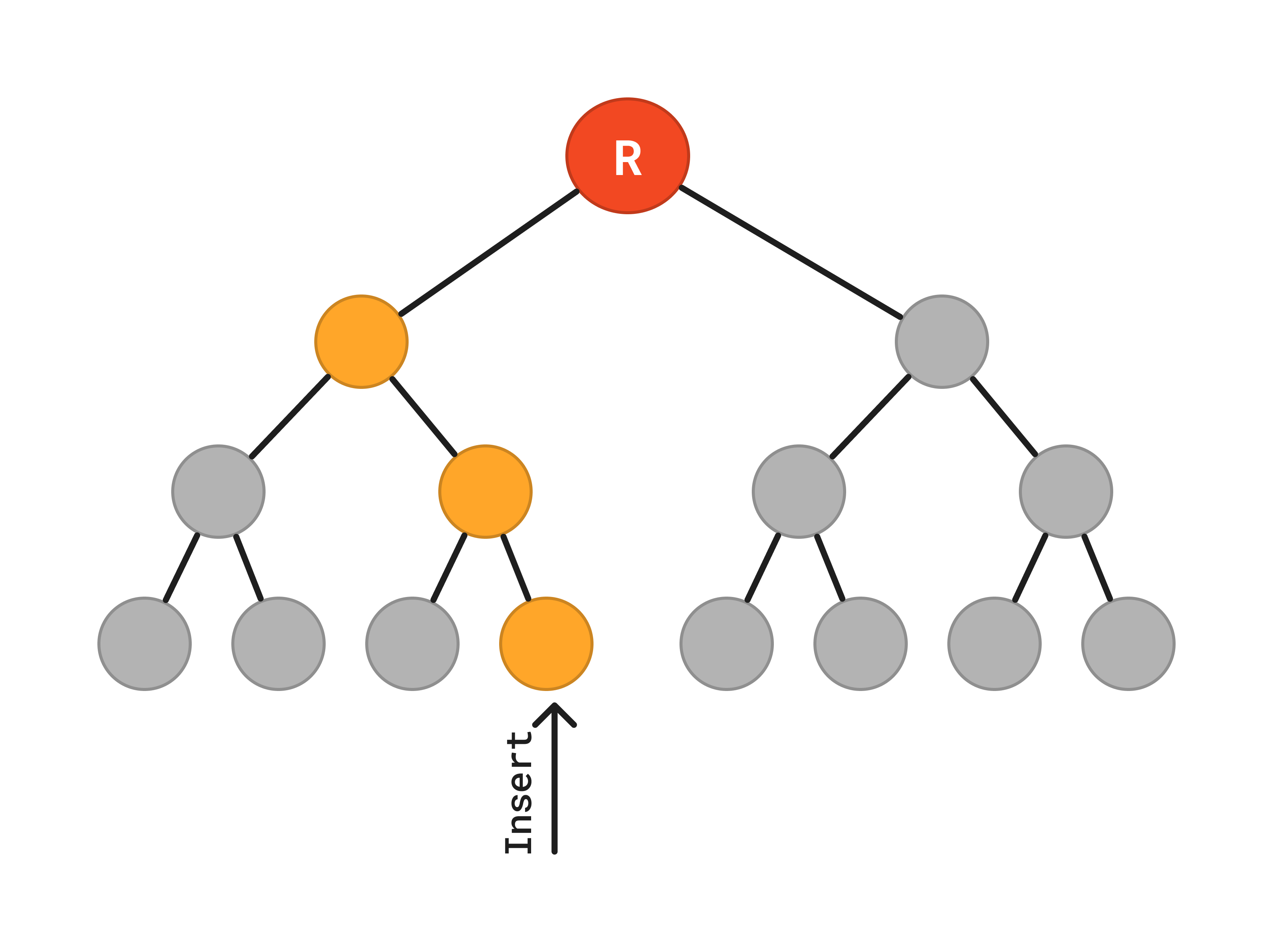}
    \caption{Merkle tree: Inserting a single data block at the leaf node alters the root.}
    \label{fig:tree}
\end{figure}

For integrity verification, only the root hash of the tree is acquired from a trusted source and matched with the root hash of data blocks received from the source. If root hashes do not match data is corrupted, otherwise not. This is because inserting a corrupted data block will lead to a different root as demonstrated in figure \ref{fig:tree}.

\subsection{Threshold Cryptosystem}
A threshold cryptosystem is a cryptosystem that allows a group of entities to share a secret key in such a way that a particular size of subset of this group is able to perform cryptographic operations, such as encryption, decryption, and digital signatures.

It has multiple benefits over the traditional systems, such as in terms of improved security, reduced risk of fraud, and increased fault tolerance.

If $n$ is the number of entities involved and $t$ is the minimum subset size ($t \leq n$) to perform the cryptographic operation then the system is called $(t, n)$-threshold system. It is possible to define operations like $(t, n)$-encryption schemes and $(t, n)$-signature schemes.

\subsection{Shamir's Secret Sharing} \label{sss}
Shamir's Secret Sharing (SSS) is an efficient method to divide a secret and distribute secret shares among participants of a group in such a way that the secret can only be recovered when a quorum of the group acts together to pool their share of the secret \cite{sss}. The individual shares of the secret does not give any information about the secret. SSS is an ideal $(t, n)$-threshold system.

SSS is often used to split encryption private keys into $n>1$  shares in order to enforce security. To reconstruct a threshold, $t \leq n$ number of shares is needed.

SSS exploits the \textbf{Lagrange interpolation} which says that one can uniquely determine a polynomial of degree $k-1$ from given $k$ points on it \cite{lagrange}. Given a set of $k$ distinct points $(x_i, y_i)$, we first compute the \textbf{Lagrange basis}, $\ell_i(x)$ corresponding to each point as:

\begin{equation}
\begin{split}
    \ell_i(x) & = \frac{(x - x_0)}{(x_i - x_0)} \frac{(x - x_1)}{(x_i - x_1)} ... \frac{(x - x_{k-1})}{(x_i - x_{k-1})} \\
    & = \prod_{\substack{0 \leq m \leq (k-1) \\ m \neq i}} \frac{x - x_m}{x_i - x_m}
\end{split}
\end{equation}

Then the resulting Lagrange interpolating polynomial of degree $k-1$ is the summation:
\begin{equation}
    f(x) = \sum_{i=0}^{k-1} y_i\ell_i(x)
\end{equation}

The theory of SSS is based on this polynomial interpolation over a finite field, 
$\mathbb{F}_q$ $(q > n)$ given a set of minimum number of points. Assume that the secret is $a_0 = s$ an element of a finite field. Choose $k-1$ elements, $a_1, a_2, ...,a_{k-1}$ randomly from $\mathbb{F}_q$ and construct a $k-1$ degree polynomial as:

\begin{equation} \label{sss-lagrange-polynomial}
    f(x) = s + a_1x + a_2x^2 + ... + a_{k-1}x^{k-1}
\end{equation}

Compute $n$ points on the curve $f$ as $(x_j, y_j)$ where $j = 0, 1, 2, ...,n-1$ and $y_j = f(x_j)$. Each of the $n$ participants receives $y_i$ as their secret share.

To recover the secret $s$ from given $k$ shares, we compute the related Lagrange basis polynomials $\ell_i(x)$. But we are only interested in recovering $s = a_0 = f(0)$, so it is sufficient to only calculate $\ell_i(0)$:
\begin{equation}
    \ell_i(0) = \prod_{\substack{m = 0 \\ m \neq i}}^{k-1} \frac{x_m}{x_m - x_i}
\end{equation}

And recover the secret $s$ as:
\begin{equation}
    s = f(0) = \sum_{i=0}^{k-1} y_i\ell_i(0)
\end{equation}

\section{Implementation of Selective De-Anonymization}

For the technical implementation of SeDe, we employ primitives from the Threshold Cryptosystem and Zero-Knowledge Proofs. ZKPs are used to constrain the user to structure the transaction as expected by the SeDe compliance framework without revealing the plain data. And threshold encryption scheme to distribute the de-anonymization responsibility among multiple accountable parties.

For simplicity, we assume that there exists an application-specific key issuer, $\mathcal{I}$ that enrolls users as members of the platform by letting them have unique identity element $id$. For privacy-preserving applications $\mathcal{I}$, for example, can be a smart contract that may store identity commitments from users, which may be used during transactions to prove membership of the platform. We assume that a note $N$ has this $id$ (or any other equivalent) as one of the fields (along with other application-dependent fields) by which the user can prove its ownership of $N$.

We also assume there is some application-defined function, $\mathtt{nullify}$ to calculate the nullifier of the given note:
\begin{equation} \label{eq:nullify}
    \eta = \mathtt{nullify}(N)
\end{equation}

\subsection{Setup} \label{setup}

We use the standard mathematical operations from elliptic curve cryptography (ECC) defined over a prime field $\mathbb{F}_q$ of order $q$ to perform all of the cryptographic operations in our implementation.

We start by defining a key generation scheme used. To generate private keys for themselves, the user and revoker can simply sample random elements from the finite field $\mathbb{F}_q$ as $p_\mathcal{U}$ and $p_\mathcal{R}$ respectively and derive corresponding public keys $P_\mathcal{U} = p_\mathcal{U} \cdot G$ and $P_\mathcal{R} = p_\mathcal{R} \cdot G$, where $\cdot$ represents the scalar multiplication of a point on the elliptic curve.

For the key pair representing the set of guardians $\mathcal{G}$, we need to use a Distributed Key Generation (DKG) scheme. Each individual private key of a guardian i.e. $p_{\mathcal{G}_i}$ must be a secret share of the whole private key $p_{\mathcal{G}}$ such that the cryptographic operations can only be performed only by $p_{\mathcal{G}}$ after it is recovered by collusion of at least $t$ guardians out of $n$ guardians. This property could be achieved by utilizing the $(t,n)$-threshold cryptosystem.

We introduce a \textit{dealer} entity which will calculate and distribute the key shares $p_{\mathcal{G}_i}$ to guardians. To achieve this, we usually employ \textit{Shamir's Secret Sharing} as described in Sec. \ref{sss}. Let $f$ be a Lagrange interpolation function such that each secret share of guardian is given as:

\begin{equation}
    p_{\mathcal{G}_i} = f(x_i)
\end{equation}

where $x_i$ could be randomly chosen field elements. The generated secrets $p_{\mathcal{G}_i}$ are distributed to guardians.

To recover the key $p_{\mathcal{G}}$ by collusion of at least $t$ individual shares, each of the $t$ guardians calculate their contribution, $b_i$ by scaling their share with the corresponding Lagrange basis function:
\begin{equation}
    b_i = \ell_i(0) p_{\mathcal{G}_i}
\end{equation}

$p_{\mathcal{G}}$ can now be evaluated simply as:
\begin{equation} \label{key-recover}
    p_{\mathcal{G}} = \sum_{i = 1}^{t} b_i
\end{equation}

We exploit the evaluation in equation \ref{key-recover} later to allow for the $(t, n)$-decryption of ciphertext encrypted with public key $P_{\mathcal{G}}$.

Moreover, for any zero-knowledge proof operations, we utilize groth16 \cite{groth16} proof system with agreed-upon public parameters represented as $\beta$. With that define a prover function by the equation:
\begin{equation}
\boldsymbol{\pi} = \mathtt{Prove}(\boldsymbol{\rho}, \boldsymbol{\omega}, \beta)
\end{equation}

where $\boldsymbol{\pi}$ is proof, $\boldsymbol{\rho}$ is public inputs and $\boldsymbol{\omega}$ is private inputs.

We also define verifier by:
\begin{equation}
    y = \mathtt{Verify}(\boldsymbol{\rho}, \boldsymbol{\pi}, \beta)
\end{equation}

where $y$ is a boolean result of verification of proof.

\subsection{Construction of Transaction} \label{tx-construction}
Constructing a transaction $T$ involves including encryptions of newly created notes $N_j$ (as a result of successful $T$) in the payload for sending the transaction. Further, we need the encryption scheme, $E$ used for encryption to be an efficient scheme in the ZK circuit because we also require a ZK proof $\boldsymbol{\pi}$ from $\mathcal{U}$ that all the computations were performed correctly.

One choice for such a scheme is the El Gamal encryption scheme \cite{el-gamal}. El Gamal requires a message that needs to be encrypted to be represented as a point on the elliptic curve. Therefore, we need to encode a note $N$ to a point on the curve, $X$. There can be multiple ways for such encoding - one plausible method being through Koblitz's method \cite{koblitz}. Additionally, a single note may have many more bytes of data to represent as a single point $X$ on the curve. To tackle that one may instead encode a note to multiple such points. But for the sake of simplicity, we represent the encoding of notes $N_j$ as a single point $X_j$.

For the encryption, the user samples a random field element $r_1$ and encrypts $X_j$ using public key $P_\mathcal{R}$ of revoker. The resulting ciphertext $E_{\mathcal{R}}(X_j)$ is a pair of points given as:

\begin{equation} \label{encryption-r}
    E_{\mathcal{R}}(X_j) = (C_{\mathcal{R}}^1, C_{\mathcal{R}}^2) = (r_1 \cdot G, X_j + r_1 \cdot P_{\mathcal{R}})
\end{equation}

For accountability and security reasons as discussed in Sec. \ref{overview}, the user wraps the data with another layer of encryption, this time using guardian public key $P_{\mathcal{G}}$.  It samples another random field element $r_2$ for the purpose and performs a similar method as in equation \ref{encryption-r}. So the points $(C_{\mathcal{R}}^1, C_{\mathcal{R}}^2)$ are encrypted to get final encryption $E_{\mathcal{G}}(E_{\mathcal{R}}(X_j))$ as:

\begin{equation} \label{encryption-g}
\begin{split}
    E_{\mathcal{G}}(E_{\mathcal{R}}(X_j)) & = (E_{\mathcal{G}}(C_{\mathcal{R}}^1), E_{\mathcal{G}}(C_{\mathcal{R}}^2)) \\
    & = ((C_{\mathcal{G}}^1, C_{\mathcal{G}}^2), (C_{\mathcal{G}}^3, C_{\mathcal{G}}^4)) \\
    & = ((r_2 \cdot G, C_{\mathcal{R}}^1 + r_2 \cdot P_{\mathcal{G}}), \\
    & (r_2 \cdot G, C_{\mathcal{R}}^2 + r_2 \cdot P_{\mathcal{G}}))
\end{split}
\end{equation}

The user also generates a proof $\boldsymbol{\pi}$ proving that everything, including the encryptions, was calculated correctly. To facilitate this, the same encryption algorithm (equation \ref{encryption-r}) can be implemented in any of the ZK DSLs (Domain Specific Languages) like circom\footnote{https://docs.circom.io/} or halo2\footnote{https://zcash.github.io/halo2/}.

The public inputs, $\boldsymbol{\rho}$ for proof generation comprise the ciphertext and the public keys:
\begin{equation}
    \boldsymbol{\rho} = (E_{\mathcal{G}}(E_{\mathcal{R}}(X_j)), P_{\mathcal{R}}, P_{\mathcal{G}})
\end{equation}

The private inputs include the plaintext point and the random elements:
\begin{equation}
    \boldsymbol{\omega} = (r_1, r_2, X_j)
\end{equation}

The inputs $\boldsymbol{\rho}$ and $\boldsymbol{\omega}$ are then passed to the prover for proof generation:
\begin{equation}
    \boldsymbol{\pi} = \mathtt{Prove}(\boldsymbol{\rho}, \boldsymbol{\omega}, \beta)
\end{equation}

Both the encryption $E_{\mathcal{G}}(E_{\mathcal{R}}(X_j))$ and proof $\boldsymbol{\pi}$ are included in the transaction payload to be sent for the execution of $T$.

\subsection{Decryption Of An Illicit Transaction}

\begin{figure*}[h]
    \centering
    \includegraphics[scale=0.34]{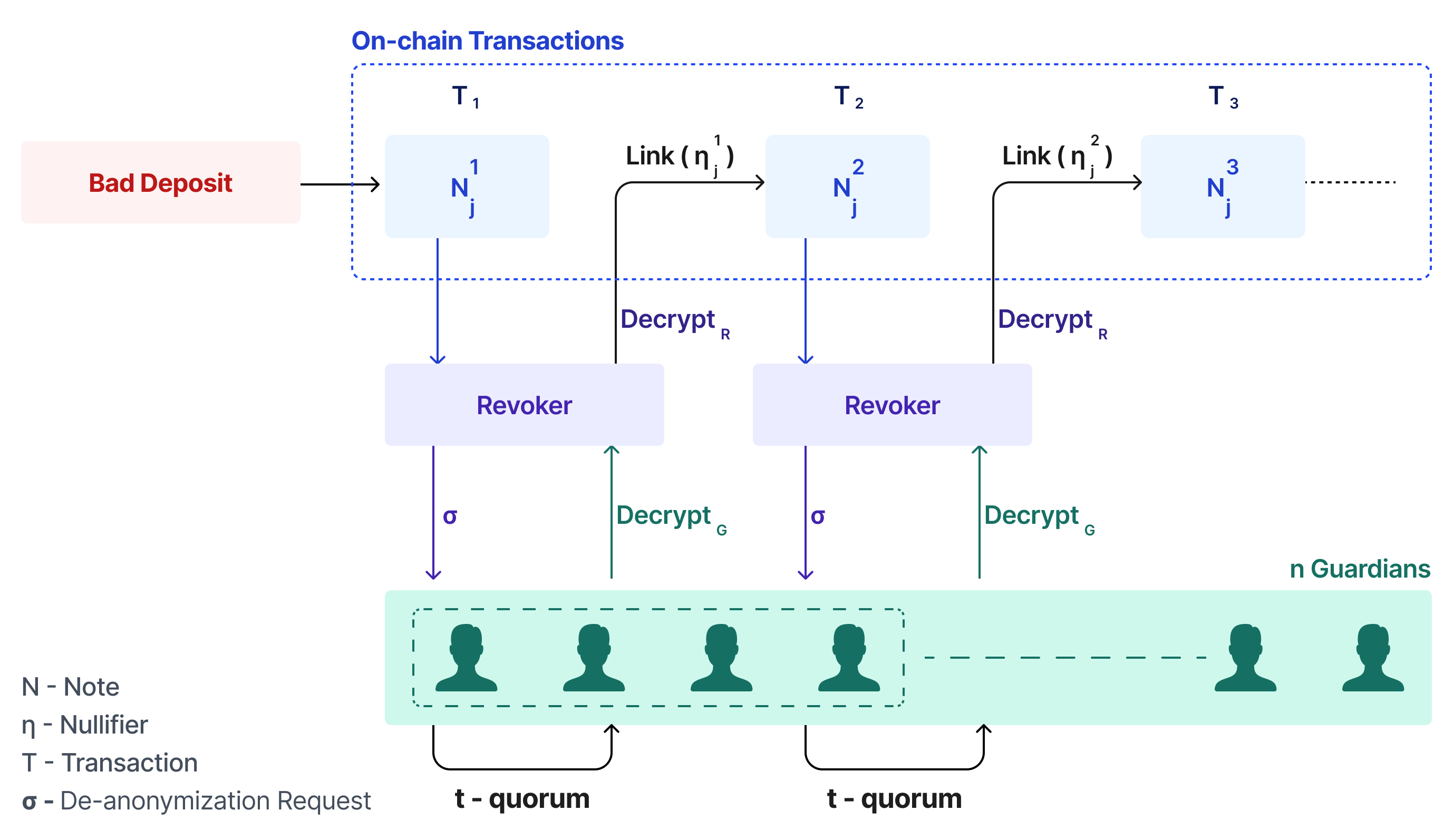}
    \caption{Linking transactions in the illicit transaction subgraph}
    \label{fig:tx-linking}
\end{figure*}

As mentioned earlier, de-anonymizing a transaction $T$ is equivalent to decrypting all the encrypted notes $N_j$ that were created by the successful execution of $T$. It means performing two decryptions corresponding to two encryptions (by $P_{\mathcal{R}}$ and $P_{\mathcal{G}}$) in order. Although the de-anonymization of transactions remains the same for any transaction in SeDe, it will usually be initiated from a deposit transaction as a starting point, as also shown in the figure \ref{fig:tx-linking}.

To request de-anonymization of a transaction $T$ ($T = T_1$ for deposit), the revoker signs $T$ to produce a signature $\sigma$ using ECDSA \cite{ecdsa}. 

\begin{equation}
    \sigma^{\prime} = \mathtt{ECDSA_{Sign}}(T, p_{\mathcal{R}})
\end{equation}

And post the de-anonymization request message, $\sigma = (T, \sigma^{\prime})$ to list of such messages in the public domain.

Guardians receive $\sigma$ and cross-check from the posted message that $T$ is already executed on the application and verify the signature as:
\begin{equation}
    o = \mathtt{ECDSA_{Verify}}(T, \sigma^{\prime}, P_{\mathcal{R}})
\end{equation}

If $o$ is $true$ then the request message is legit and guardians proceed to a decision by quorum, otherwise, reject the request.

For $t$ guardians to recover $(C_{\mathcal{R}}^1, C_{\mathcal{R}}^2)$ from the encrypted data from equation \ref{encryption-g}, each of them first calculate their contributions, $B_i$ as:
\begin{equation} \label{eq:contribution}
    B_i = b_i \cdot C_{\mathcal{G}}^1 = b_i \cdot C_{\mathcal{G}}^3
\end{equation}

These contributions can then be sent to the revoker who would then calculate the sum:

\begin{equation}
\begin{split}
    \sum_{i = 1}^{t} B_i & = \left( \sum_{i = 1}^{t} b_i \right) \cdot C_{\mathcal{G}}^1 \\
    & = \left( \sum_{i = 1}^{t} b_i \right) r_2 \cdot G \\
    & = p_{\mathcal{G}} r_2 \cdot G \\
    & = r_2 \cdot P_{\mathcal{G}}
\end{split}
\end{equation}

And then $(C_{\mathcal{R}}^1, C_{\mathcal{R}}^2)$ is recovered as:
\begin{equation}
    (C_{\mathcal{R}}^1, C_{\mathcal{R}}^2) = (C_{\mathcal{G}}^2 - \sum_{i = 1}^{t} B_i, C_{\mathcal{G}}^4 - \sum_{i = 1}^{t} B_i)
\end{equation}

To finally recover $X_j$, $\mathcal{R}$ performs simple decryption with its private key as:
\begin{equation}
    X_j = C_{\mathcal{R}}^2 - p_\mathcal{R} \cdot C_{\mathcal{R}}^1
\end{equation}

$X_j$ is simply decoded to get back plain notes $N_j$.

If the deception procedure is performed on the transaction $T = T_1$ from figure \ref{fig:tx-linking}, revoker gets the decrypted notes created by execution of $T_1$ i.e. $N^1_{j}$. Revoker would then get nullifiers, $\eta^1_j$ of $N^1_j$ using equation \ref{eq:nullify}. Now, revoker would scan the network for all transactions executed after $T_1$ where any of the calculated nullifiers $\eta^1_j$ were used as payload. Although there could be multiple such transactions, for simplicity, we take one such found transaction $T_2$. Since notes $N^1_j$ created in $T_1$ were previously identified as tainted or illicit, by definition spending any of such notes is also an illicit transaction. Therefore, $T_2$ is also identified as an illicit transaction. Revoker starts the same procedure for $T_2$ as it did with $T_1$ and links to further illicit transaction $T_3$.

In reality, multiple transactions may originate from a previous illicit transaction instead of just one. In that case, the procedure would recursively touch all such links of illicit transactions until the calculated nullifiers are not found. At that point, the currently unspent illicit notes are found. And a subgraph of illicit transactions is revealed.

\section{Example Scenario For SeDe Tracing}

Let us assume Eve has accumulated a large sum, $v_d$ of a particular asset through illicit sources e.g. DeFi protocol hacks. Since every normal transaction on blockchain is public she cannot make much use of the money. Her primary motive is to launder the assets via a private channel so that no external observer can track and link her transactions. So she proceeds to exploit a privacy protocol or application similar to as described in Sec. \ref{zk-privacy-apps}. However the application is built upon the SeDe framework, so she must adhere to the process of constructing and sending transactions.

Eve's address is already known publicly to be involved in the illicit activity (which usually is the case). To start she sends a deposit transaction, $T_1$ from her address to the application's smart contract address. This particular initial transaction can be seen publicly. Eve is now able to perform private transactions with the deposited illicit assets which cannot be linked to $T_1$ by an external observer.

Eve performs several transactions within the application as demonstrated in figure \ref{fig:example}. She performs two other consecutive transactions - one to send $v_t$ amount to one of her other accounts ($T_2$), $A_2$, and then to withdraw $v_w$ of the assets from the application ($T_3$) to any regular wallet. Neither $T_2$ nor $T_3$ can be linked to $T_1$ by an external observer. She could go on further laundering the assets ($T_4$ and $T_5$) from her accounts.

By this time, the revoker entity has already received the signal e.g. a court order or warrant, for the de-anonymization of the transaction subgraph tainted with the stolen assets. After being satisfied with the legitimacy of the event, the revoker proceeds with the steps below:

\begin{figure}[h!]
    \centering
    \includegraphics[scale=0.3]{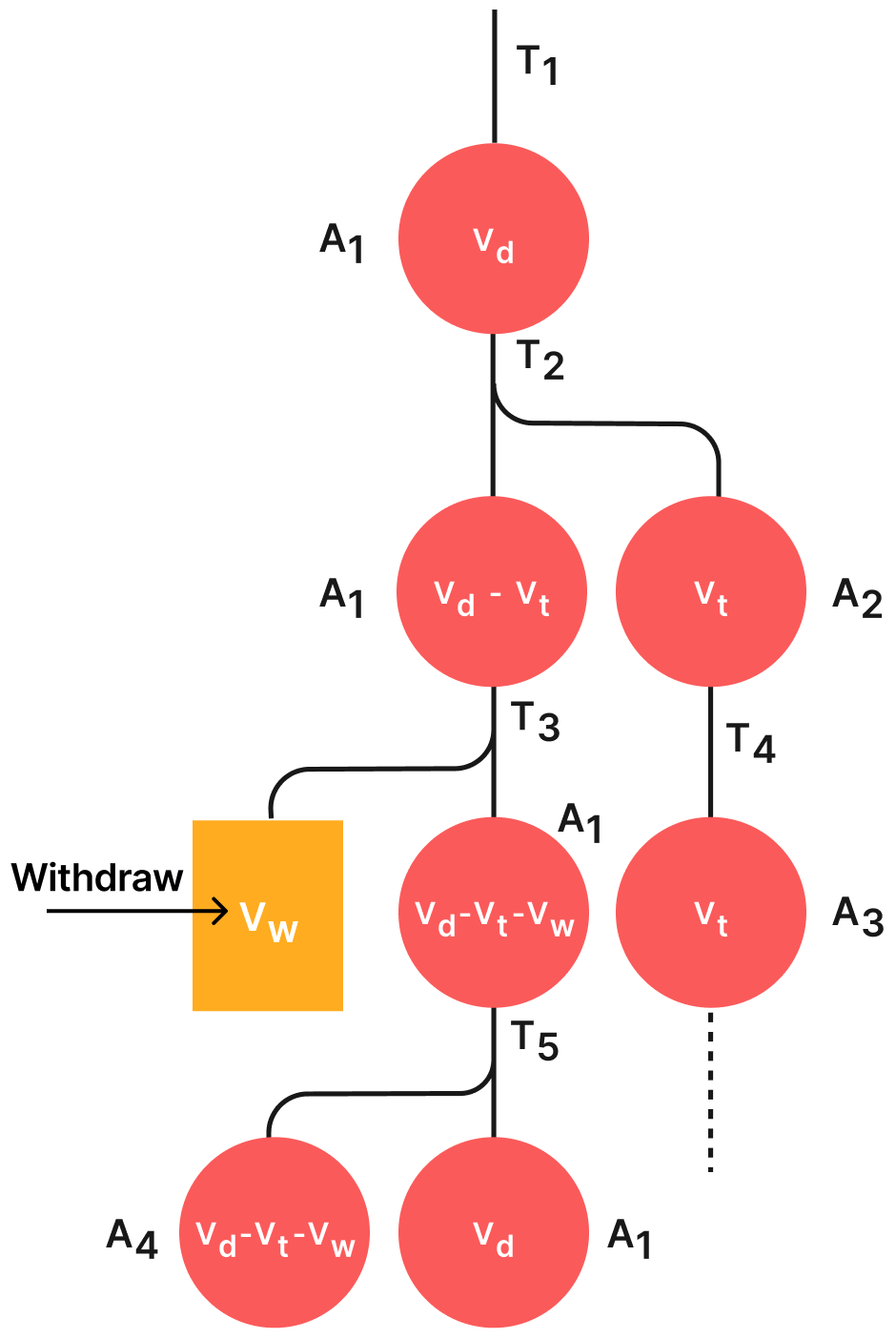}
    \caption{De-Anonymization of Eve's Transaction: Eve deposits (transaction $T_1$) $v_d$ amount of illicit money to her account $A_1$. Transfers $v_t$ to another account $A_2$ (transaction $T_2$). Then withdraws $v_w$ amount from leftover $v_d - v_t$ (transaction $T_3$). She further launders stolen assets by to other accounts $A_3$ and $A_4$ via $T_4$ and $T_5$.}
    \label{fig:example}
\end{figure}

\begin{enumerate}
    \item The revoker identifies the transaction $T_1$ and fetches the details including any encrypted data. \label{step-1}
    \item Revoker sends a verifiable message publicly that includes a signature that represents the request for de-anonymization of $T_1$ to the guardians.
    \item Guardians vote among themselves to reach a quorum for granting the de-anonymization permission by sending their secret contributions.
    \item If enough guardians agree, revoker is enabled to decrypt the transaction data by utilizing guardian contributions and its key.
    \item Being able to see plain data revoker can link deposit transaction $T_1$ to $T_2$ by calculating the relation between the expenditure of assets in $T_2$ and that which were achieved previously via $T_1$.
    \item The revoker now performs the same operation from step \ref{step-1} for de-anonymizing $T_2$ which eventually allows it to gain information such as transfer amount $v_t$ and account $A_2$ it was transferred to.
    \item Same goes for linking withdrawal transaction $T_3$ where $v_w$ was withdrawn and get the leftover balance $(v_d - v_t - v_w)$ in $A_1$.
\end{enumerate}

Any further transactions like $T_4$ and $T_5$ can be also de-anonymized recursively to finally reveal the entire subgraph of illicit transactions, $\{T_1, T_2, T_3, T_4\}$ as edges.

\section{Security}

\subsection{Trust Assumption On Dealer}

As mentioned in Sec. \ref{setup}, the setup requires an individual dealer to derive individual secret shares for guardians and distribute the same through a secure channel. This poses a critical risk if the dealer acts maliciously by not purging each of these shares after distribution. This is because a dishonest dealer with access to all of the shares easily calculates private key $p_{\mathcal{G}}$ from equation \ref{sss-lagrange-polynomial}. Although, having access to $p_{\mathcal{G}}$ does not directly allow the revelation of private transaction data, it could do so in case the revoker is also malicious and colludes with a malicious dealer. Moreover, it might defeat the purpose of guardians in certain scenarios.

To avoid the problem discussed above we need a method to remove trust assumptions on the dealer for handling secrets. Such a method can be implemented via \textit{Nested Shamir Secret Sharing} as described in \cite{nested-sss}. It presents a method for "decentralized" method for key generation such that the secret variable ($p_{\mathcal{G}}$ in our context), $a_0 = s$ from equation \ref{sss-lagrange-polynomial} cannot be seen by any party including the dealer. However, the key-generation setup requires multiple interactions among the dealer and secret recipients.

\subsection{Collusion Among Or Bribing Guardians}
A malicious actor with large social or financial capital may try to influence the decision result of guardians e.g. by bribing if the set of guardians is static and public. In that case, the majority of guardians may refuse to de-anonymize a known illicit transaction.

Chances of such attacks can be diminished by having a dynamic and/or anonymous guardian set.

\begin{itemize}
    \item \textbf{Dynamic Guardian Set}: A static set of guardians may get corrupted over time. To mitigate this one could have a dynamic set of guardians instead, in which the set of guardians changes after an epoch. This can be achieved via a dynamic proactive secret sharing (DPSS) scheme as described in \cite{dpss}. An adversary cannot prevent the de-anonymization of some illicit transactions as long as it is not able to corrupt at least $t$ guardians in every epoch.
    \item \textbf{Anonymous Guardians}: The chances of collusion among the guardians or collusion of guardians with revoker can be diminished by introducing anonymity for guardians. In such a scenario, the revoker and even guardians do not possess any information that may aid in identifying an individual or entity as a guardian. The technique devised in \cite{papr} devises a novel method for achieving the same. In our context, the method lets the user randomly and provably shuffle a list of commitments to the individual public keys of guardians and post it to the key issuer $\mathcal{I}$. $\mathcal{I}$ verifies the correctness of the random shuffle of keys and randomly selects indices in the list to assign guardians to that user. In the process, $\mathcal{I}$ gets no information about the selected subset of guardians.
\end{itemize}

\section{Improvements}

\subsection{Eliminating Un-necessary Repeated Quorums Among Guardians}
Many times the situation may require the de-anonymization of a chain of multiple linked transactions rather than a single transaction. In such cases, it is quite evident that any next transaction in the chain, linked to a previous illicit transaction, is also illicit and should be de-anonymized. The same goes for other transactions further in that chain. The de-anonymization decision of any such transaction in the subgraph of these illicit transactions should be implicit. Only the first transaction, the origin of illicit activity, should require spending time for the decision-making process.

As evident from the figure \ref{fig:tx-linking}, the current implementation implies repeated quorums for each de-anonymization request. Only the de-anonymization of $T_1$ should require any significant decision-making time. The de-anonymization decision of $T_2$ or $T_3$ could be implicit if they're known to be linked with $T_1$.

To speed up the process of decision-making the revoker, after de-anonymization of $T_1$, may prove to guardians that $T_2$ is linked to $T_1$. To do so, the receiver can generate a ZK proof, $\pi_d$ that proves to the guardians that the notes created in transaction $T_1$ were spent in the transaction $T_2$. Let's say revoker has information of notes $N^1_i$, with commitments, $c^1_i$ created in $T_1$. Consequently, it also has access to related nullifiers $\eta^1_i$ that were revealed in $T_2$. Revoker uses a ZK prover $\mathtt{Prove}_d$ to generate $\boldsymbol{\pi}_d$. $\mathtt{Prove}_d$ takes public inputs $\boldsymbol{\rho}_d$ as:
\begin{equation}
    \boldsymbol{\rho}_d \equiv (c^1_i, \eta^1_i)
\end{equation}

And private inputs as:
\begin{equation}
    \boldsymbol{\omega}_d \equiv (N^1_i)
\end{equation}

So,
\begin{equation}
    \boldsymbol{\pi}_d = \mathtt{Prove}_d(\boldsymbol{\rho}_d, \boldsymbol{\omega}_d, \beta)
\end{equation}

$\boldsymbol{\pi}_d$ proves a statement $\mathcal{S}_d$ defined as:
\begin{equation}
\begin{split}
    \mathcal{S}_d & \equiv \text{Knowledge of } \boldsymbol{\omega}_d \colon \\
    & c^1_i = H(N^1_i) \land \eta^1_i = \mathtt{nullify}(N^1_i)
\end{split}
\end{equation}

Then every individual guardian just verifies the $\boldsymbol{\pi}_d$ with a verifier $\mathtt{Verify}_d$:
\begin{equation}
    y = \mathtt{Verify}_d(\boldsymbol{\rho}_d, \boldsymbol{\pi}_d, \beta)
\end{equation}

If $y$ is $true$, then the guardians are convinced that $T_2$ is indeed linked to $T_1$ and can immediately provide its contribution in the next cycle of de-anonymization of $T_2$. This process can also be automated.

\subsection{More Efficient Encryption Mechanism}
To enforce accountability of de-anonymization, the construction of a transaction in Sec. \ref{tx-construction} requires double encryption of notes data as inspired by the technique mentioned in \cite{mutal}. With our chosen encryption scheme, encrypting one message point on the curve 
yields, in effect, four points (or three points if the random element is the same for encryption of two points in the second round of encryption) as seen in equation \ref{encryption-g}. Moreover, to get the final ciphertext of a single point, a total of three encryptions need to be performed - one for encryption by the key $P_{\mathcal{R}}$ and two more by key $P_{\mathcal{G}}$ for encryption of two points resulted from previous encryption.

This makes it a bit inefficient for target smart contract applications it is intended for, due to multiple reasons - a bigger payload for sending transactions, more gas cost, more calculations at the client side, and also a greater number of constraints in the ZK circuit.

The same effect of accountability can be achieved more efficiently by modifying the method of encryption of the same data using the same encryption scheme. Instead of two individual encryptions, user only encrypts data using the public key $Q = (P_{\mathcal{R}} + P_{\mathcal{G}})$ to get ciphertext points $E_Q(X) = (C_1, C_2) = (r \cdot G, X + r \cdot Q)$ where $r$ is randomly chosen element from field. Granting the de-anonymization then is equivalent to guardians sending their contributions $B_i$ (equation \ref{eq:contribution}) to the revoker. And revoker is enabled to decrypt the ciphertext as:

\begin{equation}
    \begin{split}
        X = C_2 - \left( p_{\mathcal{R}} \cdot C_1 + \sum^{k}_{i=1} B_i \right)
    \end{split}
\end{equation}

It achieves the same purpose as the revoker cannot perform the decryption without access to at least $k$ contributions ($B_i$) from $k$ guardians.

\section{Discussion}

\subsection{Guardian Selection}
The selection criteria of individuals for playing the role of one of the guardians can be flexible and may depend on the nature of the application or protocol. We lay out some of the possible criteria for privacy-preserving applications on blockchain.

\begin{itemize}
    \item \textit{Combining regulatory authorities and neutral gatekeepers:} One solution, as suggested in \cite{a16z} may involve making some regulatory body assign the role of the guardian(s) but also ensuring neutral gatekeeper entities through reputation. The gatekeepers would resist de-anonymization without a valid warrant or order.
    \item \textit{Selection via voting:} The protocols or applications on blockchain often have Decentralized Autonomous Organisations (DAOs) where decision is made by their members through voting. The selection guardians can be selected through voting.
    \item \textit{Selection among enrolled users:} Another idea mentioned in \cite{mutal} is the selection of the guardians among the enrolled users of the platform themselves. A verifiable random number (VRF) generator may facilitate such selection criteria for randomly choosing a set of users to take the role of guardians in each epoch.
\end{itemize}

\subsection{Guardian Incentive Compatibility}
The incentive for guardians to behave honestly and allocate availability also largely depends on the application that adopts it. However, some possible factors to support such incentives may include the following.

\begin{enumerate}
    \item \textit{Monetary:} Each guardian can receive a monthly allowance for taking active responsibility for the role.
    \item \textit{Reputation:} The guardian role can be granted to individuals or entities in the related ecosystem with high social status at stake to act maliciously.
    \item \textit{Trust:} If guardians are entities with a major public trust like individuals from the government, abusing the role can cause a loss of trust.
\end{enumerate}

\section{Conclusion}
The dilemma of privacy protection and regulatory compliance in blockchain applications deters users from adopting blockchain technology. Users should be able to conduct private transactions while remaining compliant and avoid associating with malicious actors. This paper proposes a framework for balancing user privacy and compliance. The proposed method in this paper aims to protect the users' privacy and their funds from getting mixed with those from various illicit sources by de-anonymizing only the illicit portion of the whole transaction graph. Furthermore, honest users remain unaffected by such a process, e.g. they are not required to provide proof of innocence.

\end{document}